# Towards a human eye behavior model by applying Data Mining Techniques on Gaze Information from IEC

Denis PALLEZ
CNRS UMR 6070 - I3S
University of Nice
Nice – France
**denis.pallez
@unice.fr**

Laurent BRISSON
Institut Telecom
Telecom Bretagne
CNRS UMR 2872 TAMCIC
**laurent.brisson@enst-
bretagne.fr**

Thierry BACCINO
LPEQ Lab
University of Nice
Nice – France
**thierry.baccino
@unice.fr**

**ABSTRACT**
In this paper, we firstly present what is Interactive Evolutionary Computation (IEC) and rapidly how we have combined this artificial intelligence technique with an eye-tracker for visual optimization. Next, in order to correctly parameterize our application, we present results from applying data mining techniques on gaze information coming from experiments conducted on about 80 human individuals.

## 1. INTRODUCTION

(Jaimes, Gatica-Perez et al. 2007) presents Human-Centered Computing (HCC) as an emergent field resulting from a convergence of multiple disciplines such as computer science, sociology, psychology, cognitive science… All these disciplines are concerned with understanding humans and with the design of computational devices and interfaces. However, this field is not just about the interaction, the interface or the design process but it is concerned with knowledge, people, technology and everything that ties them together. As said by the authors, Interactive Evolutionary Computation (IEC) more recently known as *human computation* (Ahn, Ginosar et al. 2006) and presented in section 2 is in the scope of HCC in the sense that humans have a central position. "Although HCC and human computation approach computing from two different perspectives, they both try to maximize the synergy between human abilities and computing resources. Work in human computation can therefore be of significant importance to HCC. On one hand, data collected through human computation systems can be valuable for developing machine-learning models. On the other hand, it can help us to better understand human behaviour and abilities, again of direct use in HCC algorithm development and system design."

In this article, we address the problem of understanding human behaviour in a particular context: the human being has to optimize a certain problem; ocular behaviour is collected with the help of an eye-tracker and analyzed with data mining techniques. We present results from this studies.

Firstly, we present in the next section interactive evolutionary computation and related works. In section 3, we present our E-TEA algorithm and the Java application that have already been proposed in order to minimize user's fatigue during interactive evaluation. In section 4, we present an experiment that we have conducted with our application in order to understand behavior of human eyes movement. Then, in section 5, results and behavior models obtained by applying data mining techniques on gaze information tracked during experiments are presented and discussed (section 6). Finally, we finish by presenting future experiments and some future works.

## 2. WHAT IS IEC?

Interactive Evolutionary Computation (IEC) is an optimization technique based on evolutionary computation such as genetic algorithm, genetic programming, evolution strategy, or evolutionary programming. Evolutionary computation consider several candidate solutions to a problem called the population. Thanks to an iterative progress, this population is computationally evolved by using mechanisms inspired by biological evolution such as reproduction, mutation, recombination, natural selection or survival of the fittest (Wikipedia 2007) according to the Darwin's theory. In classical evolutionary computation, a selection operator is often a program or a mathematical expression called the *fitness function* that expresses the quality of a candidate solution. So, Interactive Evolutionary Computation is used when it is hard or impossible to formalize efficiently this function where it is therefore replaced by a human user. A large survey of more than 250 papers can be obtained in (Takagi 2001), but the generally accepted first work on IEC is Dawkins (Dawkins 1986), who studied the evolution of creatures called "biomorphs" by selecting them manually. A very good example to better understand the interest of IEC could be photofit building (Takagi and Kishi 1999). In that case, there is no mathematical function which could specify how much a photofit is interesting; only the witness can subjectively tell whether proposed photofits are similar or not to the person he had seen before.

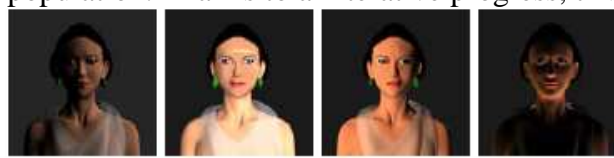

**Figure 1**: Using IEC for 3D CG lighting (Aoki and Takagi 1997)

Subsequently, much work was done in the area of computer graphics: for instance using IEC for optimizing lighting conditions for a given impression (Aoki and Takagi 1997) (cf. Figure 1), applied to fashion design (Kim and Cho 2000), or transforming drawing sketches into 3D models represented by superquadric functions and implicit surfaces, and evolving them by using divergence operators (bending, twisting, shearing, tapering) to modify the input drawing in order to converge to more satisfactory 3D pieces (Nishino, Takagi et al. 2002). We can also mention work in combining human interactions with an artificial ant, applied to non-photorealistic rendering (Semet, O'Reilly et al. 2004). Another use of IEC involves a human patient using a PDA on which an IEC is launched to define best parameter values for cochlear implants (Bourgeois-Republique, Valigiani et al. 2005). First results show that patients using PDAs obtain a better parameterization than previously through lengthy interaction with a doctor. Following the same idea of using other human senses for human interaction, we can also mention the optimization of coffee blends (Herdy 1997) by using evolution strategies.

As mentioned before, IEC is used when a fitness function is difficult and sometimes impossible to formalize. Human-Based Genetic Algorithms (HBGA) go further by allowing evolutionary computation where a good representation of individuals is hard or impossible to find (Cheng and Kosorukoff 2004), for instance they can be used in storytelling or in development of marketing slogans. To prove the usefulness of such techniques, the authors changed the classical One-Max optimization problem into an interactive one by interpreting the individuals (strings of bits – 0 or 1) as colors to be interactively presented and manipulated.

Characteristics of IEC are *inconsistencies* of individuals fitness values given by the user, *slowness* of the evolutionary computation due to the interactivity, and *fatigue* of the user due to the obligation to evaluate manually all the individuals of each generation (Takagi 2001; Semet 2002). For instance, the user is often asked to give a mark to each individual or to select the most promising individuals: it still requires active time consuming participation during the interaction. The number of individuals of a classical IEC is about 20 (the maximum that can be represented on the screen), and about the same for the number of generations.

However, some tricks are used to overcome those limits, e.g., trying to accelerate the convergence of IEC by showing the fitness landscape mapped in 2D or 3D, and by asking the user to determine where the IEC should search for a better optimum (Hayashida and Takagi 2002). Other work tries to predict fitness values of new individuals based on previous subjective evaluation. This can be done either by constructing and approaching the subjective fitness function of the user by using genetic programming (Costelloe and Ryan 2004) or neural networks, or also with Support Vector Machine (Llorà and Sastry 2005; Llorà, Sastry et al. 2006). In the latter case, inconsistent responses can also be detected thanks to graph based modeling.

Nonetheless, previous work is mostly algorithmic-oriented and not really user-oriented, which seems to be the future domain for IEC (Takagi 2001; Parmee 2007). That's why we have presented in (Pallez, Collard et al. 2007) a new technique, totally domain independent called E-TEA (Eye-Tracking Evolutionary Algorithm), to minimize this fatigue by combining an IEC and an untraditional input device. This device allows capturing user's gaze (where the user is looking on a monitor). This is possible by using eye-tracking systems such as Tobii™ which are totally non-intrusive for users. Thus, we ensure there is no need for explicit user action (choosing and clicking the most promising individual, valuating all the solutions etc.) during the evaluation process of the IEC; he just has to watch various solutions on the screen and to tell when he has finished evaluating/looking. The E-TEA algorithm then has to determine automatically which solution is better amongst presented solutions by combining gaze parameters obtained by a Tobii™. This is the work we address, applying data mining techniques on data collected during an experiment we have conducted (cf. §4).

## 3. THE EYE-TRACKING EVOLUTIONARY ALGORITHM (E-TEA)

### 3.1 What is an eye-tracking system?

An eye-tracking system consists of following the eye's motions while a user watches a screen on which something is presented. It pinpoints in real time the position where the eye is looking, with the help of one video camera focusing on a reflected infrared ray sent to the user's cornea (cf. Figure 2). This device coupled with a computer regularly samples the space position of the eye and the pupil diameter. This latter parameter lets us know the cognitive intensity of the user: the more the user is concentrated on looking at something, the smaller the diameter is (Just and Carpenter 1993). Nowadays, eye-tracking systems are very useful because they can analyze in real time what a user is focused on without any effort and in a completely non-restrictive manner. In fact, the user does not know he is being observed by a machine. With such equipment, one can finally capture when, how much time, and with which cognitive intensity a screen area is looked at.

### 3.2 How to use an eye-tracker in IEC for minimizing user's fatigue?

#### 3.2.1 The E-TEA Algorithm

A new evolutionary algorithm called Eye-Tracking Evolutionary Algorithm (E-TEA) has been proposed in (Pallez, Collard et al. 2007). It is based on a classical evolutionary algorithm eventually using breeding, selection, mutation and so on in order to evolve computationally a population of candidate solutions to a problem:

1. generate initial population randomly;
2. present the population to the user;
3. let the user watch the candidate solutions;
4. compute how much time, how many times and with which cognitive intensity the presented solutions are looked at thanks to an eye-tracker;
5. combine previously obtained parameters and compute a fitness value or a rank for each solution;
6. select the most promising solutions thanks to the computed fitness value or rank
7. make crossover and mutation
8. return to step 2 until no further good solutions are found

Thus, the user just has to watch the screen and says when he has finished watching/evaluating. There is no need for the user to mark each solution, nor to explicitly choose the best or the most promising one. This will save considerable time and the user will be capable to evaluate more solutions; consequently there will be more evaluated generations. We estimate we can double the number of evaluated screens. The main difficulty is to determine how to combine different parameters captured by the eye-tracker (step 5 of the algorithm) in order to define a computable fitness or rank.

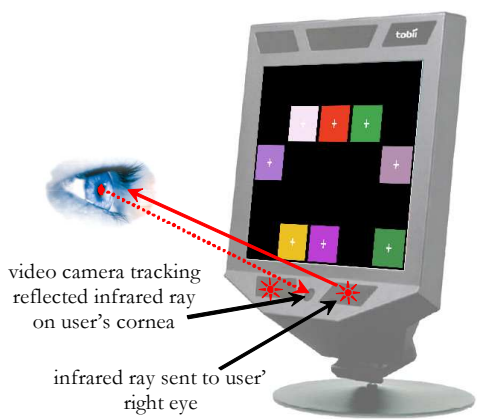

**Figure 2: How works an eye-tracker like Tobii™ 1750 ?**

The presented algorithm is not domain-dependent. However, we need to choose a specific domain to make experiments.

*3.2.2 Application to the Interactive One-Max Optimization Problem*

Our optimization problem is borrowed from (Cheng and Kosorukoff 2004) where the One-Max problem is considered as an interactive optimization problem in order to compare Interactive Genetic Algorithm (IGA) and Human-Based Genetic Algorithm (HBGA). Recall that the classical One-Max optimization problem consists in maximizing the number of 1s in a string of bits (0 or 1) only in using evolving operators (selection, mutation, crossover…). It is the simplest optimization problem and it is used here in order to parameterize our system. Basically, it consists in choosing the clearest color amongst presented colors on a screen.

*3.2.3 Our application*

We developed an application in Java 1.6 based on the Evolutionary Computation in Java library (ECJ)[1]. Solutions are represented by a string of 24 bits, 8 bits each for red, green and blue. As we capture eye motion, the screen presents only 8 zones (one solution per zone) and no individual in the center of the screen as shown in Figure 4. We avoid presenting solutions in the center because eyes are naturally attracted to the center. Also, if the user wants to compare two solutions that are diametrically opposite, eyes are obliged to cross the center. Consequently, the number of transitions for the center will increase considerably and will disrupt the estimated fitness of the solution which could be in the center. Moreover, when the application is launched, we present a screen composed of a cross in the center (cf. Figure 3a) in order to captivate the user's gaze in the center where no candidate solutions will be presented. When the gaze is concentrated on the cross (cf. Figure 3b), the next screen composed of colors is presented (cf. Figure 4). But, just before this screen of colors is presented to the user, a reference's value of the pupil diameter is computed and stored.

[1] http://www.cs.gmu.edu/~eclab/projects/ecj/

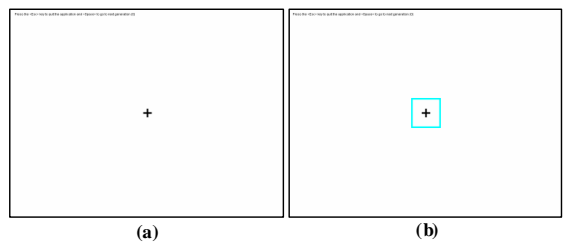

**Figure 3: Screen for fixation (a), and screen after the user had fixed the cross (b)**

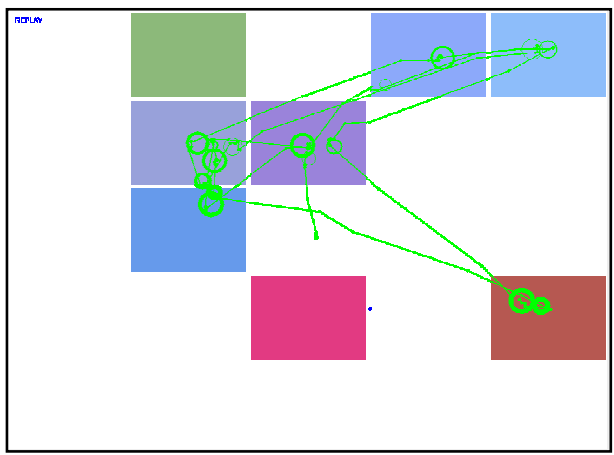

**Figure 4: Gaze information graphically represented**

When the user estimates he has finished watching solutions, we ask him to press the keyboard's space bar. When done, we detect whether the user was watching a solution. If it is the case, the solution is marked as "selected".

The issue in this algorithm is to compute either fitness or rank value for each color (that is to say for each candidate solution) from gaze information. That is why we have conducted an experiment from which we tend to determine user's behaviors.

## 4. EXPERIMENTS

### 4.1 Conditions

First of all, we have previously presented that the main goal of our research is to combine an evolutionary algorithm with an eye-tracker with the aim that an end-user and a computer collaboratively and rapidly converge to a solution satisfying the user. In fact, the user visually evaluates solutions of a problem and the computer tries to interpret user's interest for each solution. Next, the computer has to produce new solutions taking into account previous evaluation results; this is the task of evolutionary computation by using crossover, mutation and so on… By this way, we hope the new solutions will be better than previous ones. However, in this article we chose to randomly design and to randomly present solutions in order to have a better sampling of the solutions space.

When a new subject (experimenter) wants to participate, we ask him to read the following instructions: "The experiment is made up of a set of tries. Each try will proceed in two phases (Phase 1 and Phase 2). The experiment begins by the calibration of the device (the eye-tracker). All over the experiment, we recommend not to move the head. During the calibration, a blue circle is presented; fix it. Phase 1 named 'cross fixation': A white cross is presented in the center of the screen. Fix this cross to go next screen (when correctly fixed a red rectangle will surround the cross). Phase 2 named 'evaluation': Several colored squares will be presented simultaneously. Detect *color* that seems to be *lighter*. Once you think you have finished, press the space bar without looking at it to go next screen (next try)."

Phase 1 and Phase 2 are both illustrated in Figure 3 and Figure 4.

It is important to know that colors are randomly selected and set on the screen (two colors may be exactly the same or visually the same). The experiment is finished when the subject is tired (no constraint was given to subjects). Moreover, there are no dependencies between 2 screens: in fact, the evolutionary computation algorithm is not used. Nevertheless, it will be used in the future when the application will be correctly parameterized. In the following, we present the different data used in this experiment.

## 4.2 Data

### 4.2.1 Raw data coming from the eye-tracker

Data obtained from the eye-tracker (Tobii™ 1750) each 20 millisecond for each eye are the following:

- Timestamps of data in seconds and milliseconds;
- Eye position (x and y) related the current calibration;
- Eye position (x and y);
- Distance between eye and camera of the eye-tracker;
- Pupil size in millimeter;
- Validity of eye: that is whether the eye was capture or not by the eye-tracker.

In order to simplify, we only consider the gaze position represented by center of gravity of both eyes and computed from eyes positions.

### 4.2.2 Computed data

When the subject pressed the space bar indicating that he had finished visual evaluation, our application computes and store some data in files before showing next screen. Raw data are filtered in order to delete gaze positions that are called "jerk".

#### 4.2.2.1 Fixations

As the eye-tracker capture eyes position each 20 milliseconds, we need to extract some semantic information from this gaze information: what is interesting for the subject? To answer this question, we need to compute *fixations*; that is to say: what did the subject fix during movement of his eyes? According to psychologists, a fixation last between 100 and 300 milliseconds. So fixations are computed from filtered raw data. For each fixation computed, we know the following:

- Coordinates (x,y) of subject's gaze;
- Duration in microsecond;
- Colored square corresponding to the fixation. If no colored square is attached to a fixation, it is not consider as a fixation.

In raw data, the eye-tracker has given the pupil diameter and we know that it is correlated with the subject's concentration; however, we do not know how. That is why we have computed several data relating to this pupil diameter. As a fixation last at least 100 ms, a fixation is made of 5 measures at least; and we know for each of them the size of the pupil diameter. So, the following data are stored for each fixation related to the size of the pupil diameter:

- The mean;
- The size at the beginning and at the end of the fixation;
- The value of the reference pupil that corresponds of the pupil diameter when focusing on the white cross and just before presenting the colored squares (cf. Phase1 and Phase2);
- The maximum variation of the size;
- The sum of variation of the size.

Gaze path and fixations are graphically represented in Figure 4: path is symbolized by lines connecting measures each 20 milliseconds; circle symbolized fixations. Greater the radius is, longer the fixation is. The thickness of the circle is correlated to the variation of the pupil diameter.

Once all fixations are computed from filtered raw data, new data for each candidate solution (colored square) are computed. Unfortunately, fixations were not stored in files. In the future, if we need information related to fixations, we have to compute them again.

*4.2.2.2 Stored data*

As shown in Figure 4, there are several fixations for one candidate solution. Thus, we have to compute new data from fixations for each colored square (screen region or candidate solution). So, data that we had really stored are the following:

- subject's number that have participated to the experience;
- screen number evaluated by the subject;
- elements of the color model (in our case, it is Red, Green and Blue values);
- The number (called *Trans*) of transition towards the square region representing the color;
- The rank (*TransRank*) of the previous value compared with the other values of the screen;
- The sum of transition's number (*TransPop*) for all the candidate solutions of the screen;
- The relative transition's number (*TransNorm = Trans/TransPop*);
- The time (*Time*) the user has focused on a colored square;
- The rank (*TimeRank*) of the *Time* value compared with the other values of the screen;
- The sum of focused time *(TimePop)* for all the candidate solutions of the screen;
- The mean of the pupil diameter (*MeanDP*) and its relative rank *(MeanDPRank)*;
- The relative time focused on screen (*TimeNorm = Time/TimePop);*
- The reference value (*RefDP*) of the pupil diameter;
- The cognitive pupil diameter *(CognitiveDP = MeanDP-RefDP)*;
- The maximum variation of the pupil diameter and its relative rank (*MaxDPVarRank*);
- The sum of variation of the pupil diameter and its relative rank (*SumDPVarRank*);
- A Boolean value (*Selected*) representing whether the color has been fixed / "selected" just before going to next screen;
- 3 objective distances (*$M_1$, $M_2$, $M_S$*) and their relative rank;
- Positions of candidate solutions on the screen (between 0 and 1): (x,y) of upper left corner and (x,y) of bottom right corner.

Three distances for an objective fitness have been proposed in (Cheng and Kosorukoff 2004):

**(1)** $$M_1(R,G,B) = R + G + B$$

**(2)** $$M_2(R,G,B) = 255 \times \sqrt{3} - \sqrt{(255-R)^2 + (255-G)^2 + (255-B)^2}$$

**(3)** $$M_S(R,G,B) = \min(R,G,B)$$

However, the first distance has been replaced by another one that better respect the human being color model:

**(4)** $M_1(R, G, B) = 0.299R + 0.587G + 0.114B$

## 5. RESULTS

During one week and a half, 81 subjects have evaluated 7350 screens composed each by 8 colored squares. In this section we present four steps of the data mining process: analysis of physiological and behavioral data, data preparation, modeling and models evaluation.

### 5.1 Data analysis

Three data are important in order to build a predictive model to classify solutions presented on a screen: pupil diameter, time used on colored squares, and number of transitions on these squares.

Figure 5 shows the mean reference value of pupil diameter by subject. The reference value (measured before each screen) helps to determine the cognitive pupil diameter value when the user sees a colored square. The reference value depends on subject's physiology and brightness of the office where experiences take place. We can observe on that figure that the pupil diameter increases after subject number 50: it is explained by the fact that the experiment took place in two different office; the subject number 54 corresponds to a new office. We can see mean reference value of pupil diameter varies between 2.8mm and 5.8mm.

Figure 6 shows the mean time spent for observing one screen by subject. The time varies between 0.5s and 3.5s (if we don't consider some exceptions).

Figure 7 shows minimum, maximum and mean number of transitions on one colored square by subject. We can notice that, for each subject, some colored squares aren't watched. By mean, a subject sees 0 to 2 colored squares. The maximum value varies between 3 to 12 transitions towards a colored square: it is an exceptional behavior which underlies a doubt.

This behavior can be explained by the fact that the subject uses his parafoveal vision that is not detected by the eye-tracker.

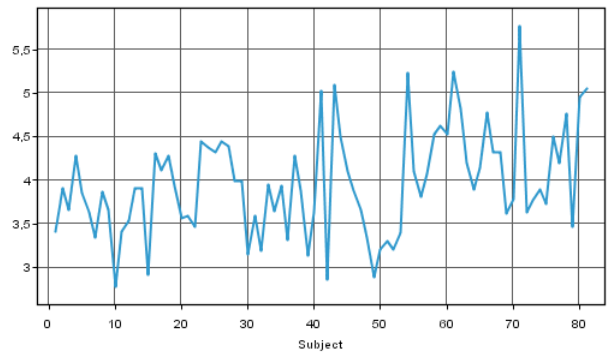

**Figure 5: Reference value of pupil diameter by subject**

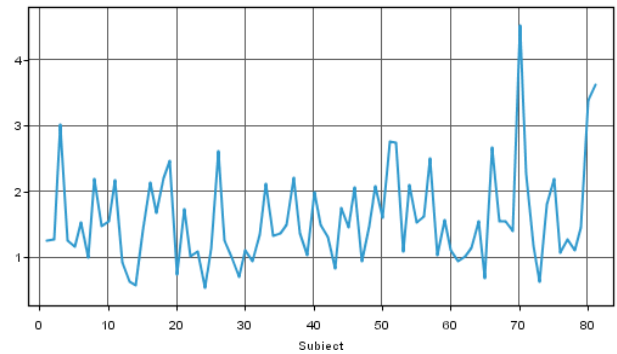

**Figure 6: Mean time used for observing one screen by subject**

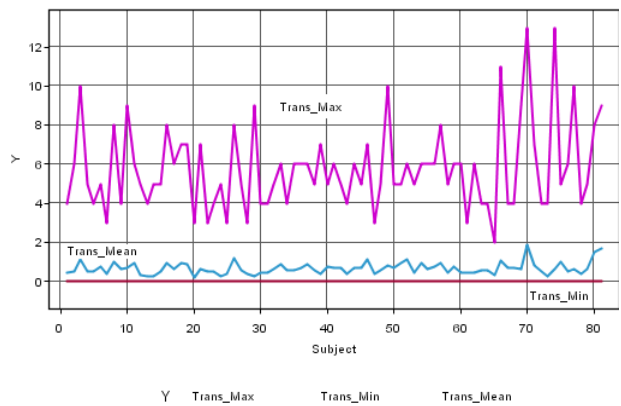

**Figure 7: Min, max and mean number of transitions on one individual by subject**

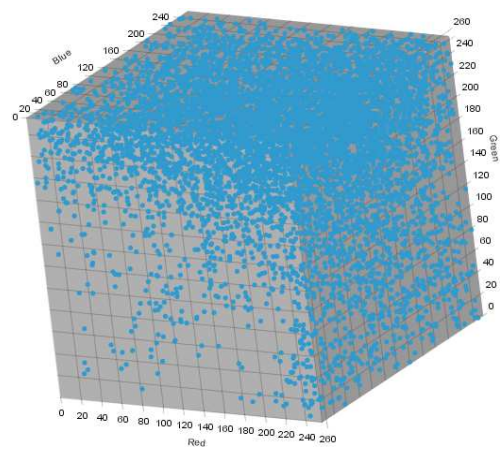

**Figure 8: Distribution of selected colors in the RGB model**

Figure 8 presents distribution of user selected colors according the RGB model. We can notice an high density of selected colored squares which maximize R,G and B values. If we consider all of the colors displayed on the screen, there is an uniform distribution.

Figure 10 shows if the subject has correctly chosen the colored square according to M1 and M2 distances. If we consider M2, users rarely made the right choice, while if we consider M1 users made the right choice once on two. Thus, we can consider M1 is the best distance measure in order to build our model.

### 5.2 Data preparation

*5.2.1 Discretization*

M1 is the distance we chose to use in order to build our model. Figure 9 shows the number of colored squares according M1 value. Since the distribution isn't uniform it is necessary to discretize this distance in order to construct some sets of values whose size is equals. This operation is necessary to avoid learning biases.

We created five sets:

- Darker: M1 ∈ [1,81[
- Dark: M1 ∈ [81,112[
- Undefined: M1∈ [112,141[
- Light: M1∈ [141,172[
- Lighter: M1∈ [172,251]

*5.2.2 Data selection*

In a first step, we create two datasets in order to try different paradigms with relative attributes or rank attributes. First set, called A, contains as predictive attributes *TransNorm, TimeNorm, CognitiveDP* and the second set, called B, contains as predictive attributes *TransRank, TimeRank, MeanDPRank, MaxDPVarRank, SumDPVarRank*. The target attribute is the discrete M1 value.

In a second step, we created two sets A' et B' which contains one more attribute: *Selected*. These datasets are available on our web webpage[2]:

---

[2] http://perso.enst-bretagne.fr/laurentbrisson/activites-recherche/osef/

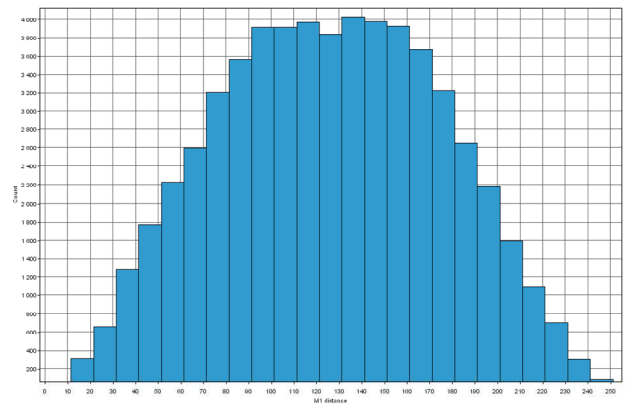

**Figure 9: Number of colored squares according M1 value**

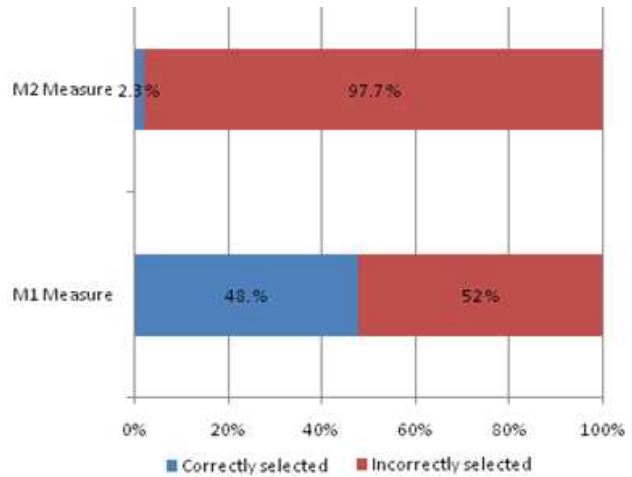

**Figure 10: Distribution of subjects choices according M1 and M2 distances**

### 5.3 Modeling

We use C5.0 algorithm, a widely used and tested decision tree algorithm successor of C4.5. Description of this algorithm is beyond the scope of this paper and the reader should refer to (Quinlan 1993).

We created two sets each one with one square on two. The first set is the learning set and we use 15 subsets for cross validation, while the second set is a validation set for evaluating models.

We generated 4 models with the same parameters for each dataset A, A', B, B'.

### 5.4 Evaluation

Figure 11 shows results of the models. We notice that the best model is the one using dataset A without *Selected* attribute. Figure 12 shows coincidence matrix for model A in order explain these results. For each row, we can observe distribution of predicted values. For example, 4760 darker colored square are correctly classified however 444 of them were classified as "Ligther". We can notice the algorithm often predicts a colored square as *"Darker"* which implies lots of errors. To solve this problem we tried cost matrix without significant results.

|  | A | A' | B | B' |
|---|---|---|---|---|
| Correctly classified | 28.81% | 28.37% | 28.02% | 28.19% |

**Figure 11: Models evaluation**

| ↓ Real / Predicted → | Darker | Dark | Undefined | Light | Lighter |
|---|---|---|---|---|---|
| Darker | 4760 | 268 | 166 | 288 | 444 |
| Dark | 4359 | 343 | 169 | 348 | 667 |
| Undefined | 3892 | 320 | 237 | 401 | 998 |
| Light | 3235 | 331 | 212 | 413 | 1739 |
| Lighter | 2269 | 285 | 231 | 307 | 2718 |

**Figure 12: Coincidence matrix for model A**

## 6. DISCUSSIONS

The Eye-Tracking Evolutionary Algorithm is a very simple but very innovative proposition that is at the intersection of two different domains: computer and cognitive sciences. This approach presents many advantages:

– First, it is the first time that an eye-tracker takes a very active part in a computer application. More traditionally, eye-tracking systems are used for analyzing human behavior when looking at an image, a text, a 3D model, a webpage, etc.
– Second, with such a combination we automate interactive evaluation of individuals with no constraints for the user. The only thing he has to do is to watch individuals and to say when he has finished. There is no explicit task imposed on the user, and thus no additional fatigue.
– Next, such material is completely non-intrusive, i.e., the user could forget that he is being observed. Interactive evaluation is as natural as possible.
– Finally, by analyzing the cognitive activity of the user, we can easily detect when the user is tired. "PERCLOS" measure is the most reliable and valid determination of a user's alertness level. PERCLOS is the percentage of eyelid closure over the pupil over time and reflects slow eyelid closures ("droops") rather than blinks. A PERCLOS drowsiness metric was established in a 1994 driving simulator study as the proportion of time in a minute that the eyes are at least 80 percent closed (Wierwille, Ellsworth et al. 1994).

Of course, each new system has its drawbacks, but they are few compared to the advantages:

– The eye-tracker can follow eyes if and only if it has been calibrated to the user. However, this takes only few seconds, and the user just has to focus on concentric moving circles.
– The other small constraint is that the user does not have total freedom of head movement. For instance, he can not look away and then resume evaluating. However, the freedom is large enough (30x16x20 cm) because of the use of two video cameras. If the signal is lost for one eye, the eye-tracker uses the other eye.

## 7. CONCLUSION AND FUTURE WORKS

In this article, we have presented a combination of a classical optimization technique represented by Interactive Evolutionary Computation and less classical device (an eye-tracker). Result is an innovative approach to minimize user's fatigue during interactive evaluation of proposed solutions to an optimization problem.

Before proving that this approach is better than others which use a classical device as a mouse, we need to correctly parameterize our application and understand human eye behavior by an experiment: ask people to detect the lightest color amongst 8 presented colors. During this experiment, data were stored and analyzed in order to find models of behavior for human eye movements. We have to improve resulting models in order to better know how to combine ocular data for computing either a fitness value or a rank value for each candidate solution. Once better models will be found, we'll have to prove that our E-TEA algorithm is better than others using a mouse. To do that, we need to conduct another experiment by evolving solutions with the help of evolutionary computation rather than presenting random solutions. Next, we also need to integrate in our application a machine learning module that will be able to predict fitness or rank value for each candidate solution as already mentioned in (Takagi 2001). Finally, with all these modifications, it will be interesting to test our approach in a real world application.

## 8. ACKNOWLEDGMENTS
We would like to thank the Institute of Technology at Nice University (http://www.iut-nice.fr), polytech school of Nice Sophia-Antipolis (http://www.polytech.unice.fr/) which let this research work be possible and persons which participated to experiment..